\documentstyle[preprint,aps,epsf]{revtex} 

\begin{document} 

\title{
Cluster Analysis for Percolation on Two Dimensional Fully Frustrated System
}  
\author{
Giancarlo Franzese
}
\address{
Dipartimento di Scienze Fisiche,
Universit\`a di Napoli, Mostra d'Oltremare Pad.19 I-80125 Napoli
Italy\\
and INFM - unit\`a di Napoli
} 

\date{\today}

\maketitle 

\begin{abstract} 
The percolation of Kandel, Ben-Av and Domany clusters for 2d fully frustrated 
Ising model is extensively studied through numerical simulations.
Critical exponents, cluster distribution and fractal dimension of
percolative cluster are given.

\end{abstract} 

\newpage

\section{Introduction}

A 2d fully frustrated (FF) Ising model is a model with Ising spins $\pm 1$
where the interactions between 
nearest neighbor spins have modulus $J>0$ and sign $\pm 1$ (ferro / 
antiferromagnetic interactions) and where the 
signs are chosen in such a way that every {\it plaquette} (i.e. the 
elementary cell of square lattice) is {\it frustrated}, i.e. every plaquette 
has an odd number of -1 interactions so that
the four spins of the plaquette cannot simultaneously satisfy all four 
interactions. In Fig.1.1 we give an example of such a deterministic interaction 
configuration.
In a plaquette of the FF model we can have only one or three 
satisfied interactions. The FF model has an analytical solution 
\cite{Villain_Forgacs} and a critical temperature at $T_c=0$. 

Since single-spin dynamics for FF suffers critical slowing down, a fast 
cluster dynamics was introduced by Kandel, Ben-Av and Domany (KBD) in 
Ref.\cite{KBD}. 

The KBD-clusters are defined by choosing stochastically on each plaquette of a 
checkerboard partition of a square lattice one 
bond configuration between the three shown in Fig.1.2.

The probability of choice depends on spins configuration on the plaquette and 
it is a function of temperature ({\it correlated site-bond percolation} 
\cite{CSK}). When there is only one satisfied interaction the zero-bond 
configuration is chosen with probability one. When three interactions are 
satisfied the zero-bond configuration is chosen with probability 
$P_0=e^{-4J/(kT)}$ (where $k$ is the Boltzmann constant and $T$ the 
absolute temperature), the bond configuration with two parallel bonds on two 
satisfied interactions is chosen with probability $P_1=1-P_0$ and the third 
bond configuration has zero probability. Two sites are in the same cluster if 
they are connected by bonds. For sake of simplicity from now on we choice 
$J/(kT)=1/T$.

Ref.\cite{KBD} has stimulated several works \cite{Ker_Cod,articoli} 
that pay attention mainly to dynamics and to number of clusters and 
clusters sizes. In \cite{articoli} numerical simulations on relatively large FF
lattice sizes (number of sites $N=60^2\div 120^2$) supported the idea that the 
KBD-clusters represent spin correlated regions (like Coniglio-Klein clusters 
\cite{CK} do in Ising model) and consequently percolation temperature $T_p$ coincides 
with critical temperature $T_c$, percolation exponents coincide with 
critical ones and KBD-clusters at $T_p$ are 2d self-avoiding walks (SAW) at 
$\theta$ point. \cite{Co_Jan}

In this paper we extensively study percolative features of KBD-clusters, 
considering very large lattice sizes ($N=100^2\div 400^2$), and give 
numerical results on critical exponents, cluster distribution and 
fractal dimension at percolation point.

\section{Critical exponents and percolation point}

We consider finite systems with increasing size ($L=100 \div 400$) with 
periodic boundary conditions.

A cluster percolates when it connects two opposed system sides. For 
every size $L$ there is a percolation temperature $T_p(L)$. With $T_p$ 
(without any argument) we mean the percolation temperature in the thermodynamic 
limit i.e. $T_p(L)\rightarrow T_p$ for $L\rightarrow \infty$.
In this limit percolating clusters are present at $T\leq T_p$ 
but not at $T>T_p$.
In Fig.2 we show typical clusters at several temperatures for a finite system 
with size $L=60$. 

For every $L$ we have studied the mean cluster size $S=\sum_s 
s^2n_s$ (where $s$ is the cluster size, $n_s$ the number of cluster of size 
$s$ per lattice site and the sum is extended over all finite clusters), the 
percolation probability $P=1-\sum_s s n_s$, the number of cluster 
$N_c=\sum_s n_s$, the number of bonds per lattice site $N_b$, the mean 
size of the largest (percolating)
cluster $S_I$ and the mean size of the second largest (percolating) 
cluster $S_{II}$. These 
quantities are shown in Fig.3 for $L=100 \div 400$. Let's note that for 
$T\rightarrow 0$ the bonds cover 50\% of lattice interactions (that is the 
random-bond percolation threshold on square lattice), 
$S_{II}$ goes to a finite value (like predicted by KBD 
\cite{KBD} and already verified in Ref.\cite{Ker_Cod}) 
and occupies almost 35\% of the lattice, and that $S_I$ occupies 
almost 65\% of the lattice. At $T=0$ 
only two clusters survive, as shown in Fig.2 e) and f). 

Now we will give numerical estimates of critical exponents that characterize 
the KBD-cluster percolation.

We know \cite{Stauffer} that in the thermodynamic 
limit the mean cluster size diverges for $T\rightarrow T_p$, the percolation 
probability goes to zero in the limit $T\rightarrow T_p^-$ and the number of 
cluster goes to zero for $T\rightarrow T_p^+$.

We assume
that near $T_p$ the {\it connectivity length} $\xi$ 
(i.e. the typical linear cluster size) diverges like $\xi\sim  
|e^{-2/T} - e^{-2/T_p}|^{-\nu}$ , the mean cluster size diverges like 
$S\sim |e^{-2/T} - e^{-2/T_p}|^{-\gamma}$, the percolation 
probability goes to zero like $P\sim |e^{-2/T} - e^{-2/T_p}|^{\beta}$ and 
the number of cluster goes to zero like $N_c\sim |e^{-2/T} - 
e^{-2/T_p}|^{2-\alpha}$. The
last relations are definitions of critical exponents $\alpha$, $\beta$, 
$\gamma$ and $\nu$.

By standard finite-size scaling considerations \cite{Stauffer} we can make the 
{\it ansatz}
\begin{equation}
S\sim L^{\gamma/\nu}f_S(|e^{-2/T} - e^{-2/T_p}| L^{1/\nu})
\label{scaling_S}
\end{equation}
\begin{equation}
P\sim L^{-\beta/\nu}f_P(|e^{-2/T} - e^{-2/T_p}| L^{1/\nu})
\label{scaling_P}
\end{equation}
and 
\begin{equation}
N_c\sim L^{(\alpha-2)/\nu}f_{N_c}(|e^{-2/T} - e^{-2/T_p}| L^{1/\nu})
\label{scaling_Nc}
\end{equation}
where $f_S(x)$, $f_P(x)$ and $f_{N_c}(x)$ are {\it universal functions}, i.e.
 independent by $L$.

Via data-collapse (see Fig.4) we estimate the parameters $e^{-2/T_p}=0.0000$, 
$\alpha=0.1$, $\beta=0.00$, $\gamma=2.00$, 
and $\nu=1.00$ with error of one unit in the last given digit . Therefore the 
scaling relation $\alpha+2\beta+\gamma=2$ and the hyperscaling relation 
$2-\alpha=\nu d$ are satisfied with good approximation.

In Tab.1 we give numerical estimates of $T_p(L)$. 
The data are obtained taking for $L=100, 200, 300, 400$ the values of $T_p(L)$ 
at which the $S$ data in a log-log plot vs. $|e^{-2/T}-e^{-2/T_p(L)}|$ follow  
two parallel straight lines (one above and one below $T_p(L)$) with slopes in 
good agreement with $\gamma=2$ and then best-fitting these values as 
$e^{-2/T_p(L)}\sim 1/L$.

\section{Fractal dimension and cluster distribution}

Let's now consider the fractal dimension $D$ of the percolating cluster. 
From the scaling invariance hypothesis \cite{Stauffer} we know that $P\sim 
\xi^{D-d}$, then we obtain $D=d-\beta/\nu$ (hyperscaling). 
In present case we have $\beta=0$, then $D=d=2$. The same result is obtained 
from the scaling relation $\beta+\gamma=D\nu$.

This is confirmed by the analysis of cluster distribution (see Fig.5). 
The scaling 
invariance hypothesis \cite{Stauffer,Donorio} gives for $T\rightarrow T_p$ and 
$s \rightarrow \infty$
\begin{equation}
n_s=s^{-\tau}f_{n_s}(|e^{-2/T} - e^{-2/T_p}| s^{\sigma})
\label{scaling_ns}
\end{equation}
with $\tau=1+d/D$, $\sigma=1/(\nu D)$ and $f_{n_s}(x)$ universal function. 

From data-collapse for $n_s$ near $T_p$ (see Fig.6.a) we obtain numerical 
estimates of parameters. The data in Fig.6.a are chosen in such a way that the 
quantity $(e^{-2/T} - e^{-2/T_p})L^{1/\nu}$ (with $e^{-2/T_p}=0$ and $\nu=1$)
is a constant with $T\simeq T_p(L)$ for every considered $L$.
The results are $\tau=2.00$ and 
$\sigma=0.50$ (with error of one unit in the last digit), that, with the 
definitions of $\tau$ and $\sigma$, give $D=2$ and $\nu=1$. On the other hand 
these values of $\tau$ and $\sigma$ satisfies the relation 
$\sigma(2-\alpha)=\tau -1$, $\sigma\beta=\tau-2$, 
$\sigma\gamma=3-\tau$.\cite{Stauffer} 

From Fig.6.a we see that the universal function $f_{n_s}(x)$ is a bell-shaped 
curve for $T\simeq T_p(L)$. For temperatures slightly below $T_p(L)$ (Fig.6.b) 
$f_{n_s}(x)$ is shifted, while for temperatures slightly above $T_p(L)$ 
(Fig.6.c) $f_{n_s}(x)$ changes dramatically its shape.

Away from $T_p(L)$ we know \cite{Stauffer} that is valid the relation 
\begin{equation}
\log n_s \sim -s^{\zeta}
\label{zeta}
\end{equation}
for $s\rightarrow \infty$, with $\zeta=1$ above $T_p(L)$ and $\zeta=1-1/d=1/2$ 
below $T_p(L)$. This relation is confirmed with reasonable approximation by our
numerical simulations, as shown in Fig.7. Let's note that, while the exponent 
$\zeta=1$ above $T_p(L)$ is good for a wide range of $s$ ($s=2000\div 8000$ 
for $L=100$), the exponent $\zeta=1/2$ below $T_p(L)$ is good for a 
smaller $s$ range ($s=2000\div 4400$ for $L=100$) since 
finite-size effect become more important below $T_p(L)$. The smaller $T$, the 
smaller $s$ range is.

A direct way to estimate the fractal dimension $D$ is given through 
its definition
\begin{equation}
s\sim R^D
\label{def_D}
\end{equation}
for $T=T_p$, with $R$ radius of gyration of the cluster of size 
$s$. We know \cite{Stauffer} that cluster dimension deviates from $D$ away from
$T_p$, becoming the Euclidean dimension $d$ below $T_p$ and a value smaller 
than $D$ above $T_p$. This is true because eq.(\ref{def_D}) is valid within the 
connectivity length $\xi$ for all temperatures, but $\xi$ goes to zero away 
from $T_p$. Unfortunately data about this relation are difficult to analyze. 
Indeed  near $T_p(L)$ 
for every finite system with $L\leq 120$ it seems that $D$ is almost $7/4=1.75$ 
(the fractal dimension of a SAW at $\theta$ point), but for larger $L$ (see 
Fig.8 and Tab.1) 
the fractal dimension $D$ grows slowly to the asymptotic value 2.

\section{Conclusions}

We have numerically investigated the KBD-cluster percolation 
problem in 2d FF Ising model. From our simulation we found that, within 
numerical errors, this 
correlated site-bond percolation satisfies scaling and hyperscaling relations 
and have, in the thermodynamical limit, a percolation temperature $T_p=0$ and 
the exponents $\alpha=0$, $\beta=0$, $\gamma=2$, $\nu=1$, $\tau=2$, 
$\sigma=1/2$, $\zeta(T>T_p(L))=1$, $\zeta(T<T_p(L))=1/2$. 
Moreover at $T_p$ clusters are compact (fractal dimension $D=2$). 
Therefore now we can correct the conclusion of Ref.\cite{articoli} 
and say that, since $T_p=T_c$ and $\nu$ is equal to spin 
correlation exponent \cite{nota}, the site connectivity length $\xi$ goes 
like the spin correlation length diverging at zero temperature.
Although the exponent $\gamma$ is different 
the coincidence between $\xi$ and correlation length is 
enough to give an efficient Monte Carlo cluster dynamics.\cite{PhyRevE}

\section*{Acknowledgments}

The author is indebted to Antonio Coniglio and to Vittorio Cataudella 
for many illuminating discussions and a careful reading of the manuscript.

The computation have been done on DECstation 3000/500 with Alpha processor 
and DECsystem 5000/200 with RISC processor.

\newpage

\begin{table}  
\caption{Numerical estimates of $T_p(L)$ and $D(L)$ for $L=60\div 400$. 
The way used to evaluate $T_p(L)$ and $D(L)$ 
give us confidence only on digit not in parentheses.
}    
\begin{tabular}{c|c c c c c c c}  
$L$ 	& 60	& 80	& 100 	& 120	& 200 	& 300 	& 400  \\ 
\hline
$T_p(L)$&0.51(7)&0.48(1)&0.45(6)&0.43(7)&0.39(2)&0.36(1)&0.34(2) \\ 
$D(L)$	& 1.7(2)& 1.7(5)& 1.7(7)& 1.7(9)& 1.8(2)& 1.8(5)&1.8(6) \\
\end{tabular}
\end{table}  

\begin{center}

\begin{figure}
\mbox{ \epsfxsize=13cm \epsffile{ 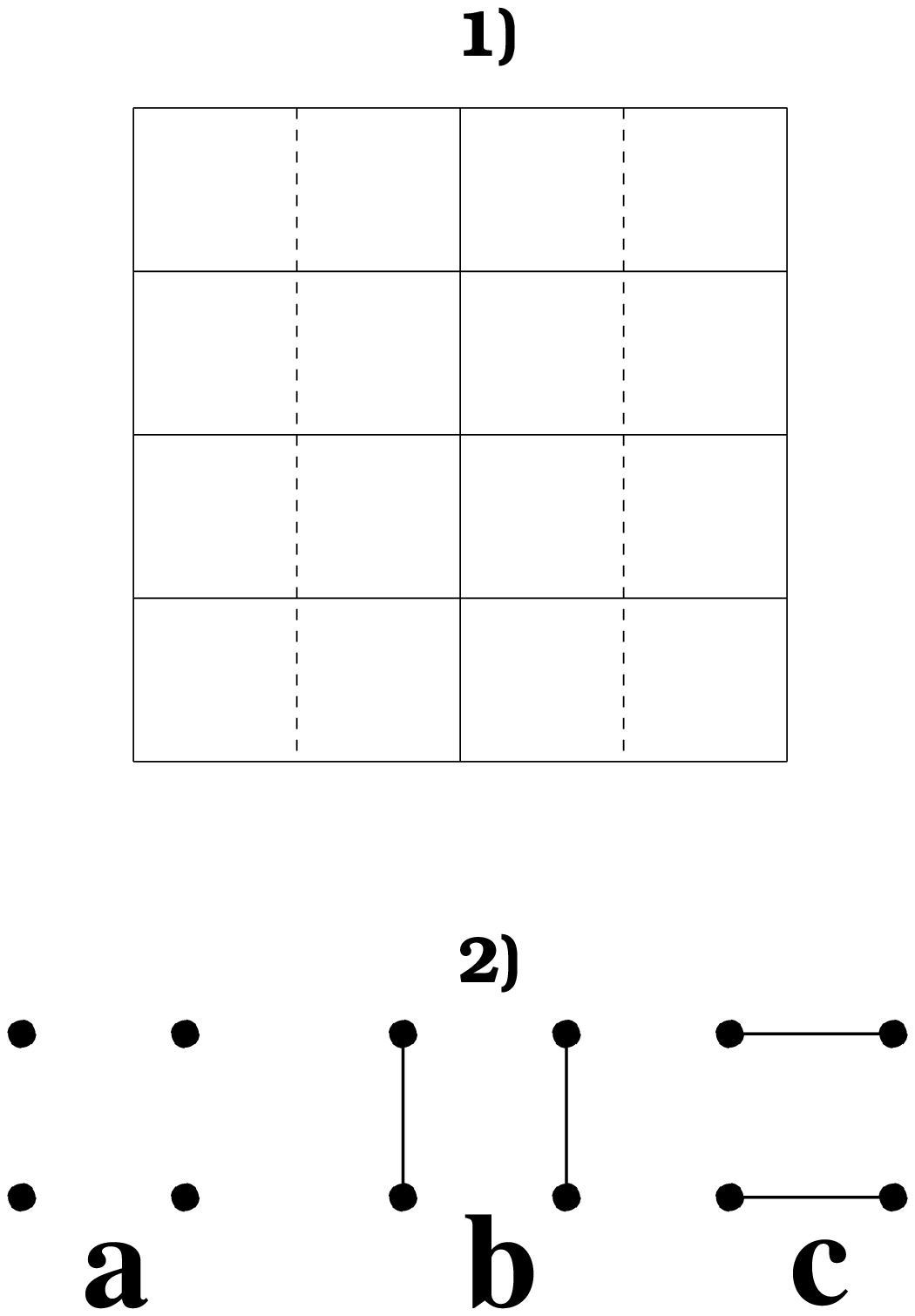 } } 
\vspace{1cm}
\caption{ 1) Example of 2d FF lattice: the spin are on the vertices;
solid lines represent ferromagnetic interactions ($+J$) and dashed lines 
antiferromagnetic interactions ($-J$). 
2) Plaquette bond configurations: a) zero bond; b) two parallel vertical bonds;
c) two parallel horizontal bonds. }
\end{figure}

\begin{figure}
{\Huge a} \mbox{ \epsfxsize=4cm \epsffile{ 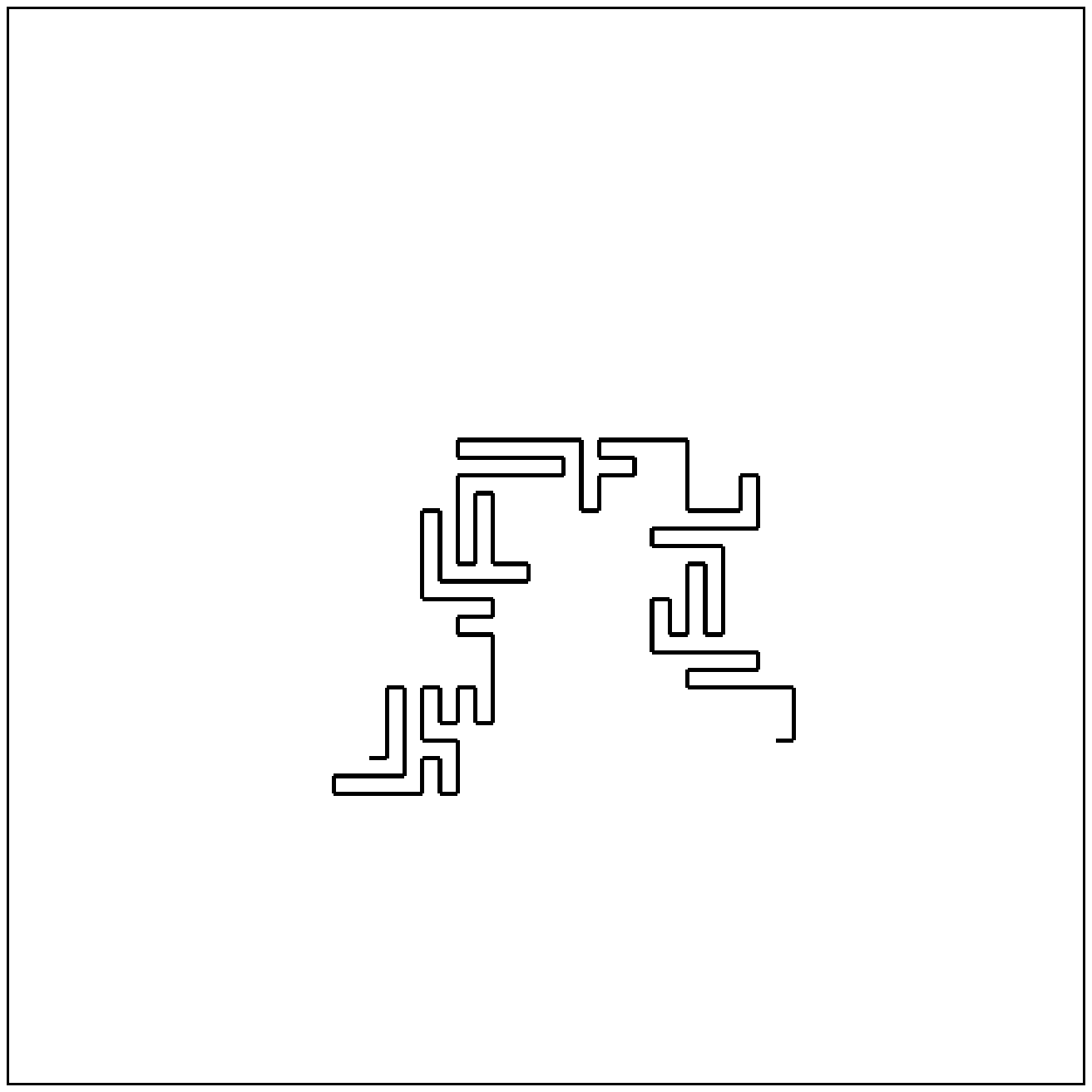 } } 
{\Huge b} \mbox{ \epsfxsize=4cm \epsffile{ 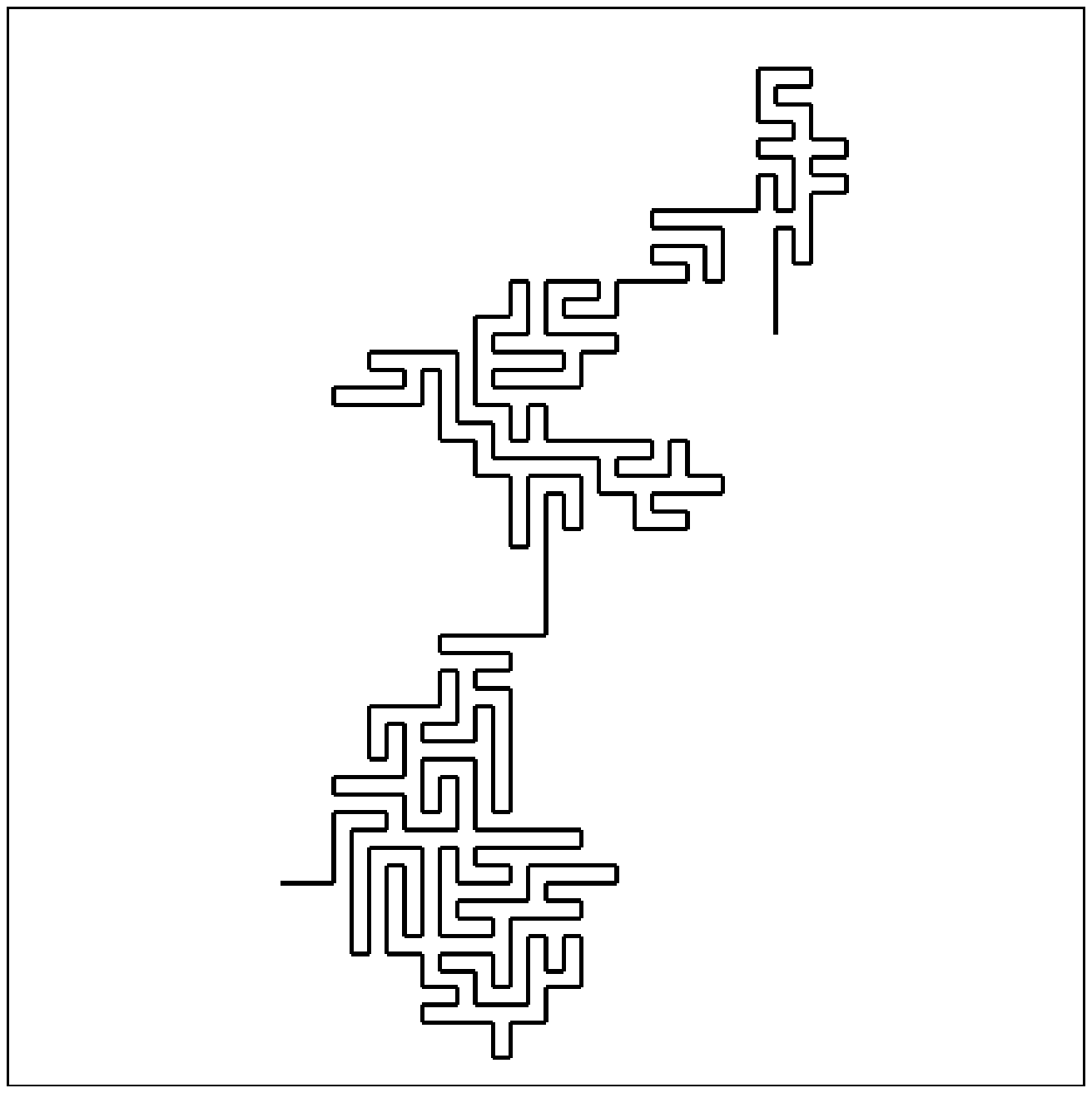 } }

\vspace{1cm}

{\Huge c} \mbox{ \epsfxsize=4cm \epsffile{ 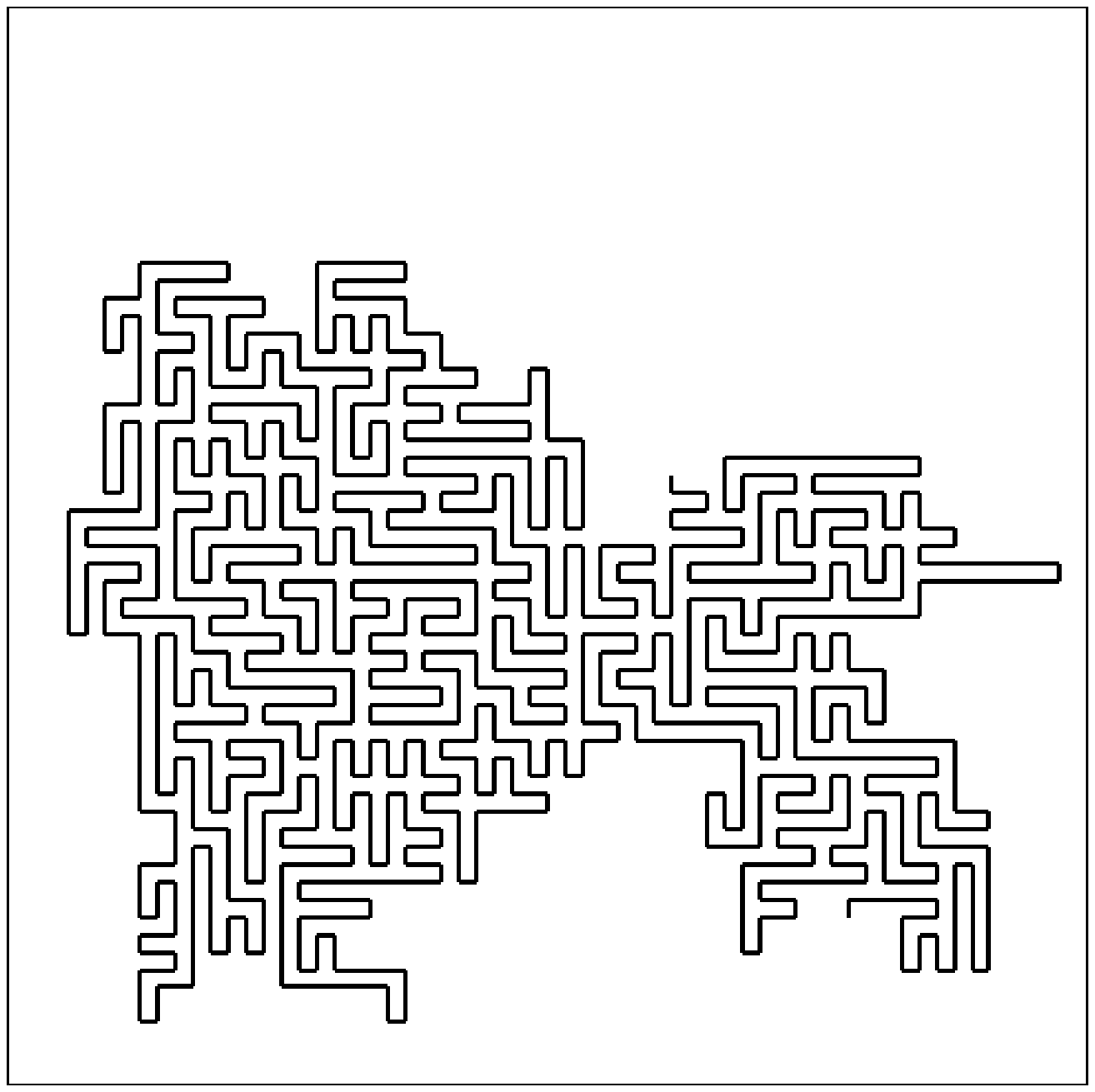 } }
{\Huge d} \mbox{ \epsfxsize=4cm \epsffile{ 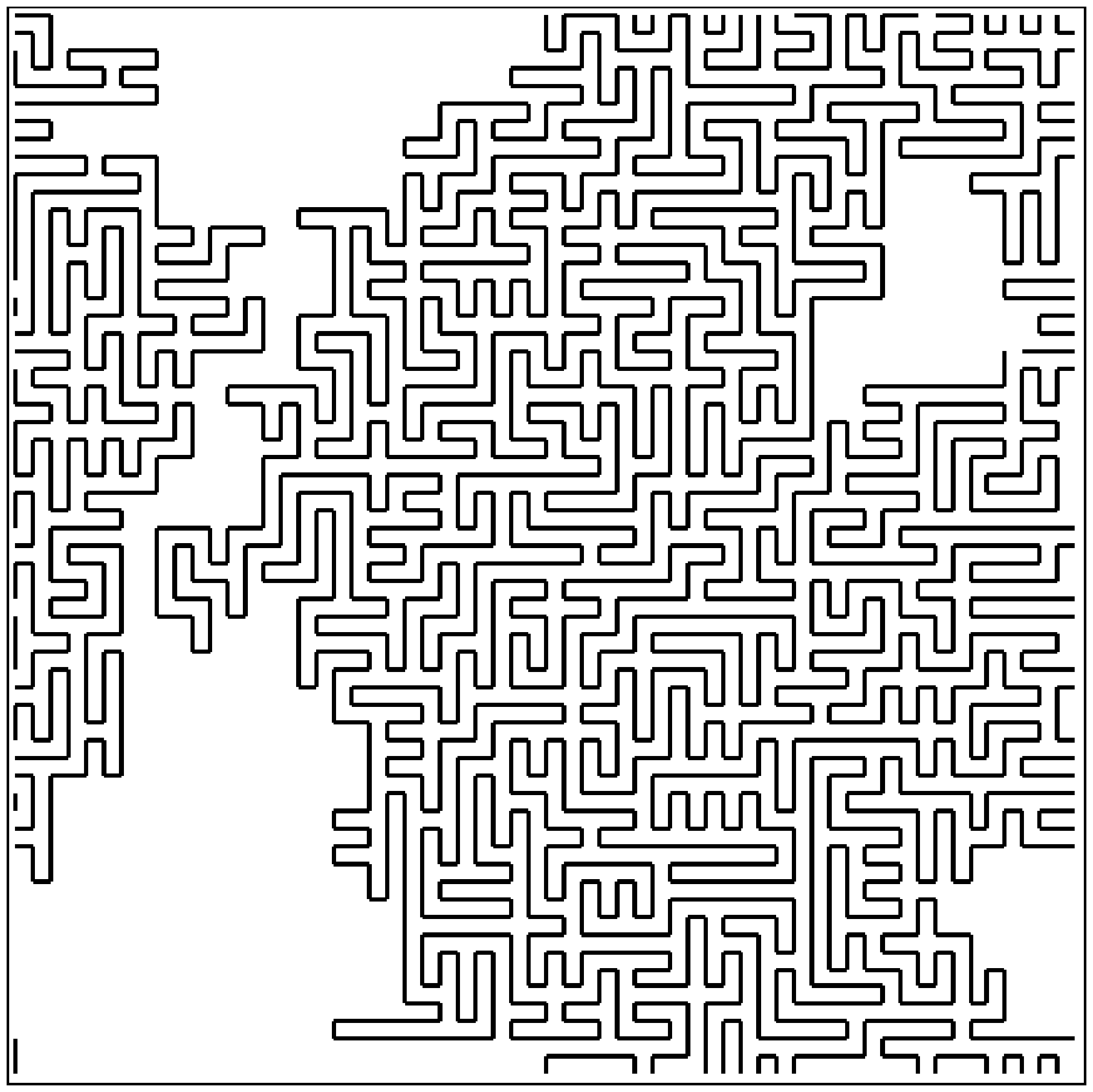 } }

\vspace{1cm}

{\Huge e} \mbox{ \epsfxsize=4cm \epsffile{ 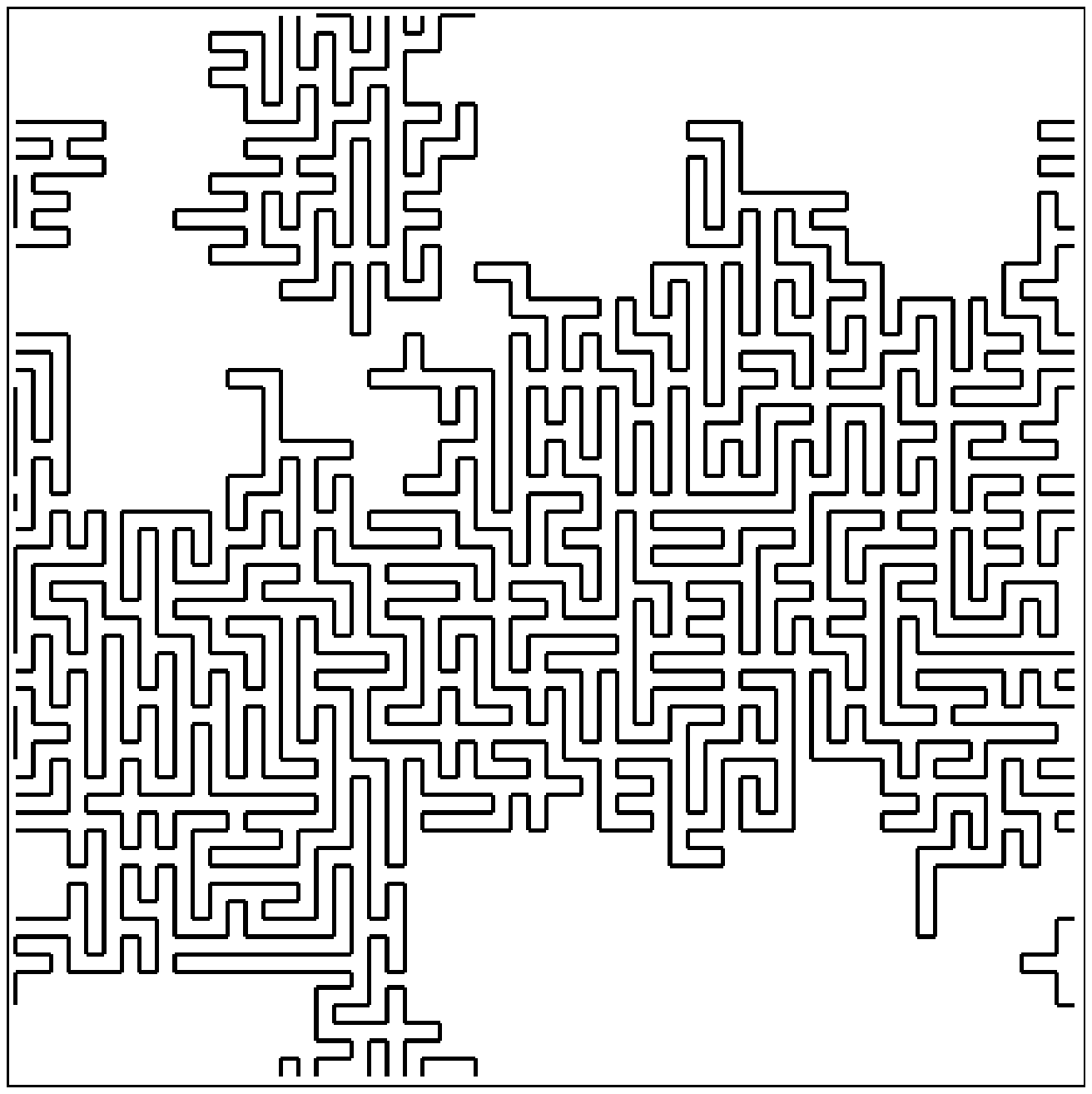 } }
{\Huge f} \mbox{ \epsfxsize=4cm \epsffile{ 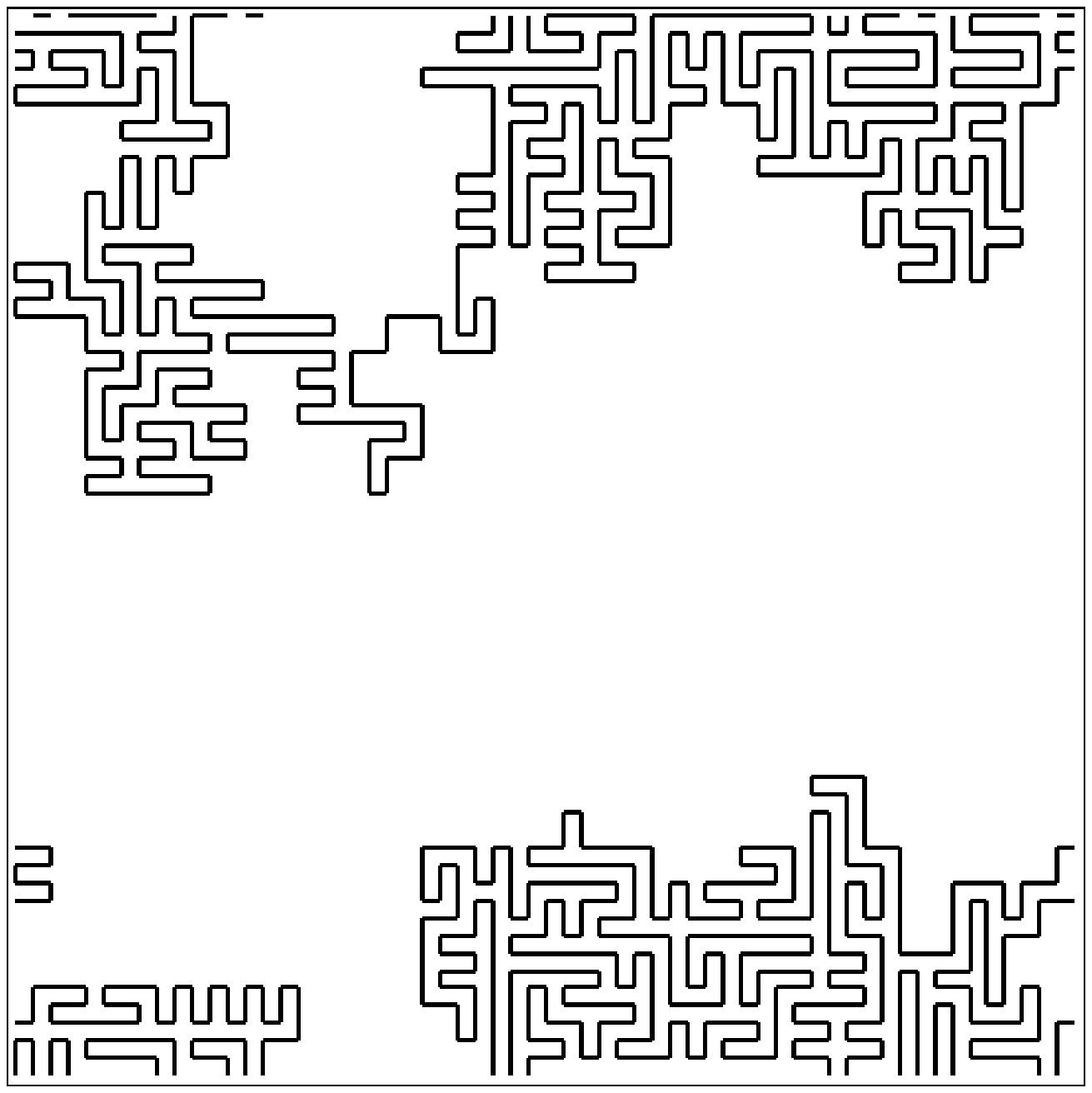 } }

\vspace{1cm}
\caption{Typical KBD-clusters on a FF lattice with size $L=60$ with 
periodic boundary conditions: a) at $T=1$; b) at $T=0.65$; c) at 
$T=0.53$ slightly above $T_p(L)\simeq 0.52$; d) at $T=0.52\simeq T_p(L)$; e) 
at $T\simeq 0$ (largest cluster); f) at $T\simeq 0$ (the second cluster).
}
\end{figure}

\begin{figure}
\mbox{ \epsfxsize=13cm \epsffile{ 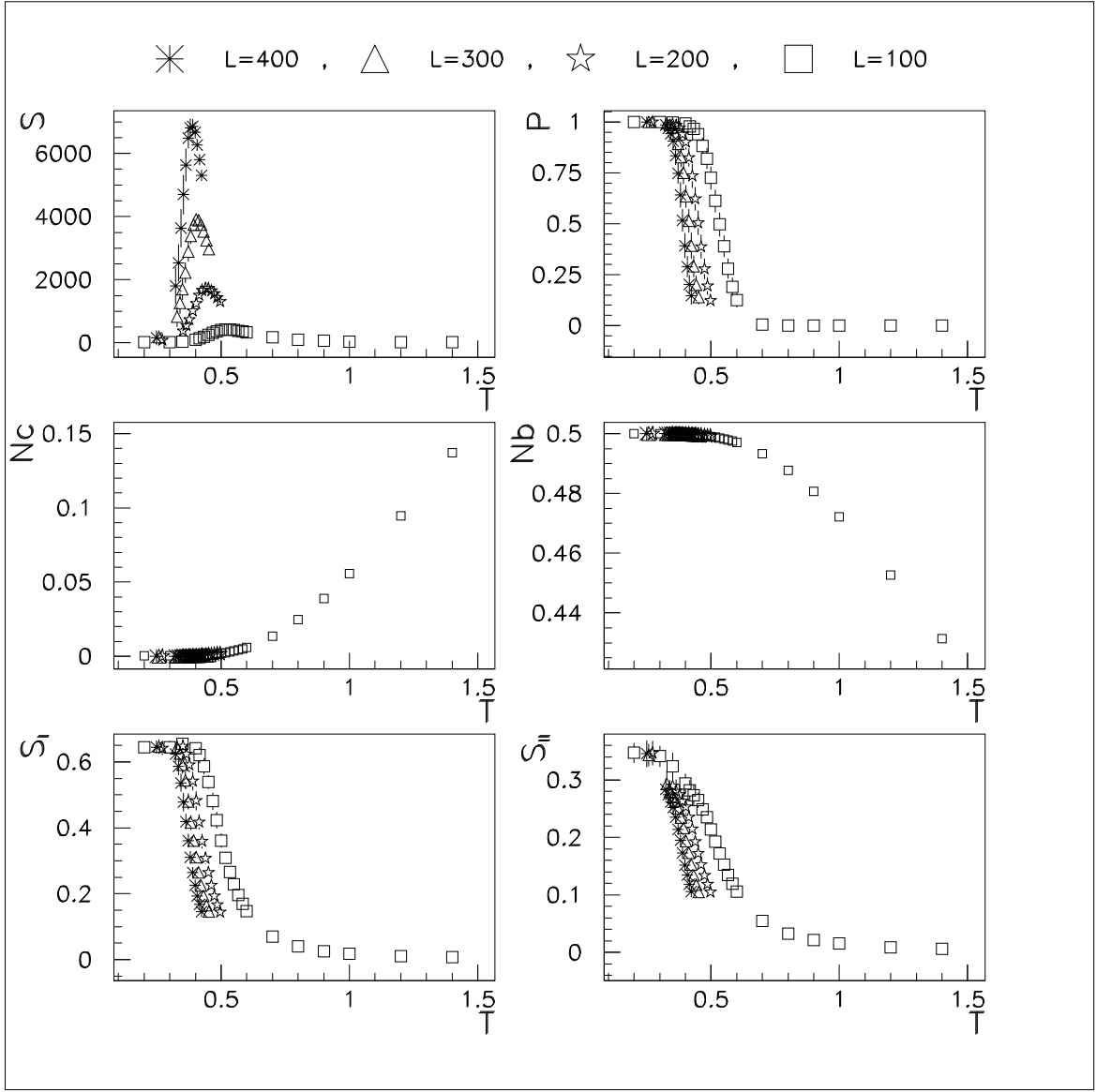 } }
\caption{Mean cluster size $S$, percolation probability $P$, number of 
cluster $N_c$, number of bonds per lattice site $N_b$, mean size of largest 
cluster $S_I$ and mean size of second largest cluster $S_{II}$ vs. temperature 
$T$ for lattice sizes $L=100,200,300,400$. The error bars (often included in 
symbols) are the statistical errors.
}
\end{figure}

\begin{figure}
\mbox{ \epsfxsize=10cm \epsffile{ 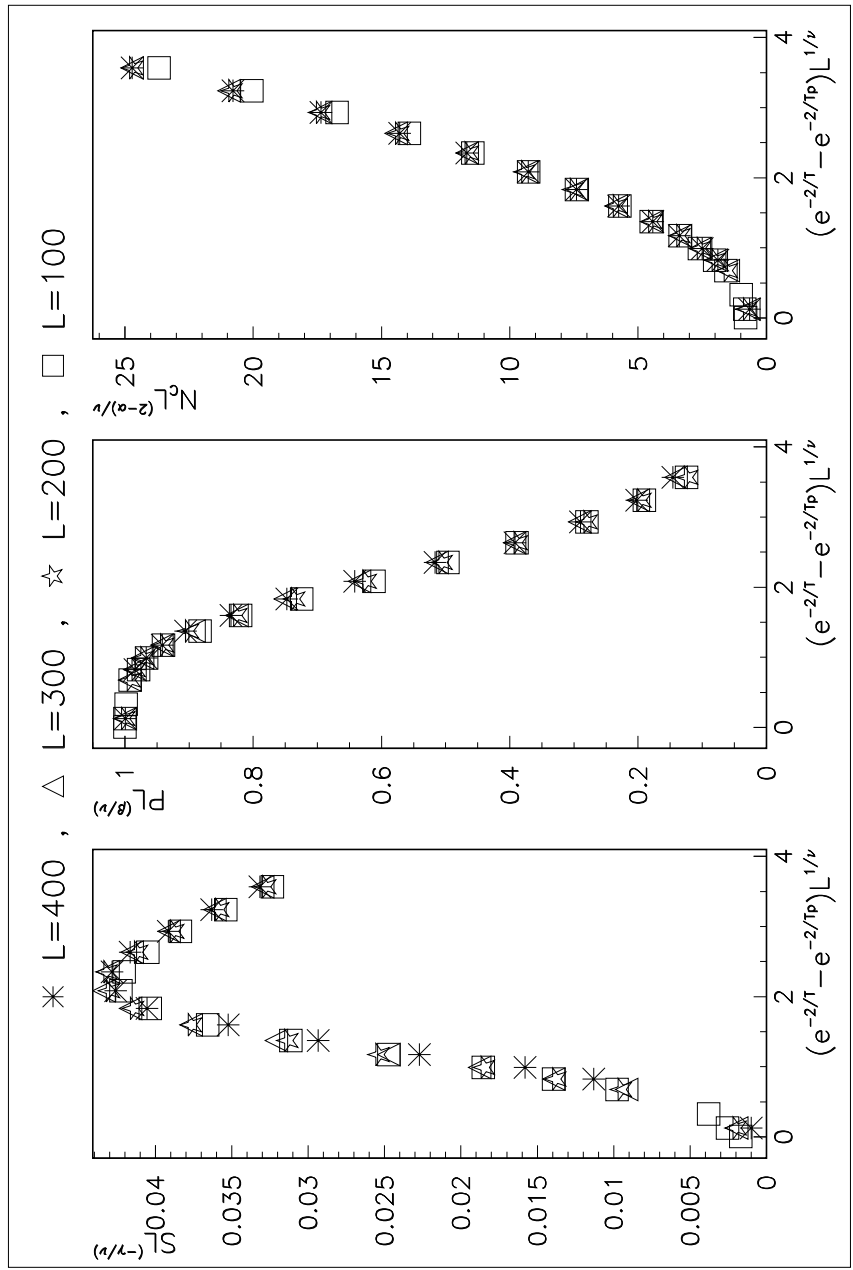 } }

\vspace{1.5cm}

\caption{Scaling for $S$, $P$ and $N_c$ following assumptions
(\ref{scaling_S}), (\ref{scaling_P}) and (\ref{scaling_Nc}) for data of 
systems with sizes $L=100,200,300,400$. The parameters 
$e^{-2/T_p}=0.0000\pm 0.0001$, 
$\alpha=0.1 \pm 0.1$, $\beta=0.00\pm 0.01$, $\gamma=2.00\pm 0.01$ and $\nu=1.00
\pm 0.01$ are such that the data for different sizes $L$ collapse on 
single curves (one for each graph). These curves are, respectively, the 
universal functions $f_S$, $f_P$ and $f_{N_c}$. The errors are 
estimated observing the range of parameters within which the data points 
collapse roughly well.
}
\end{figure}

\begin{figure}
\mbox{ \epsfxsize=13cm \epsffile{ 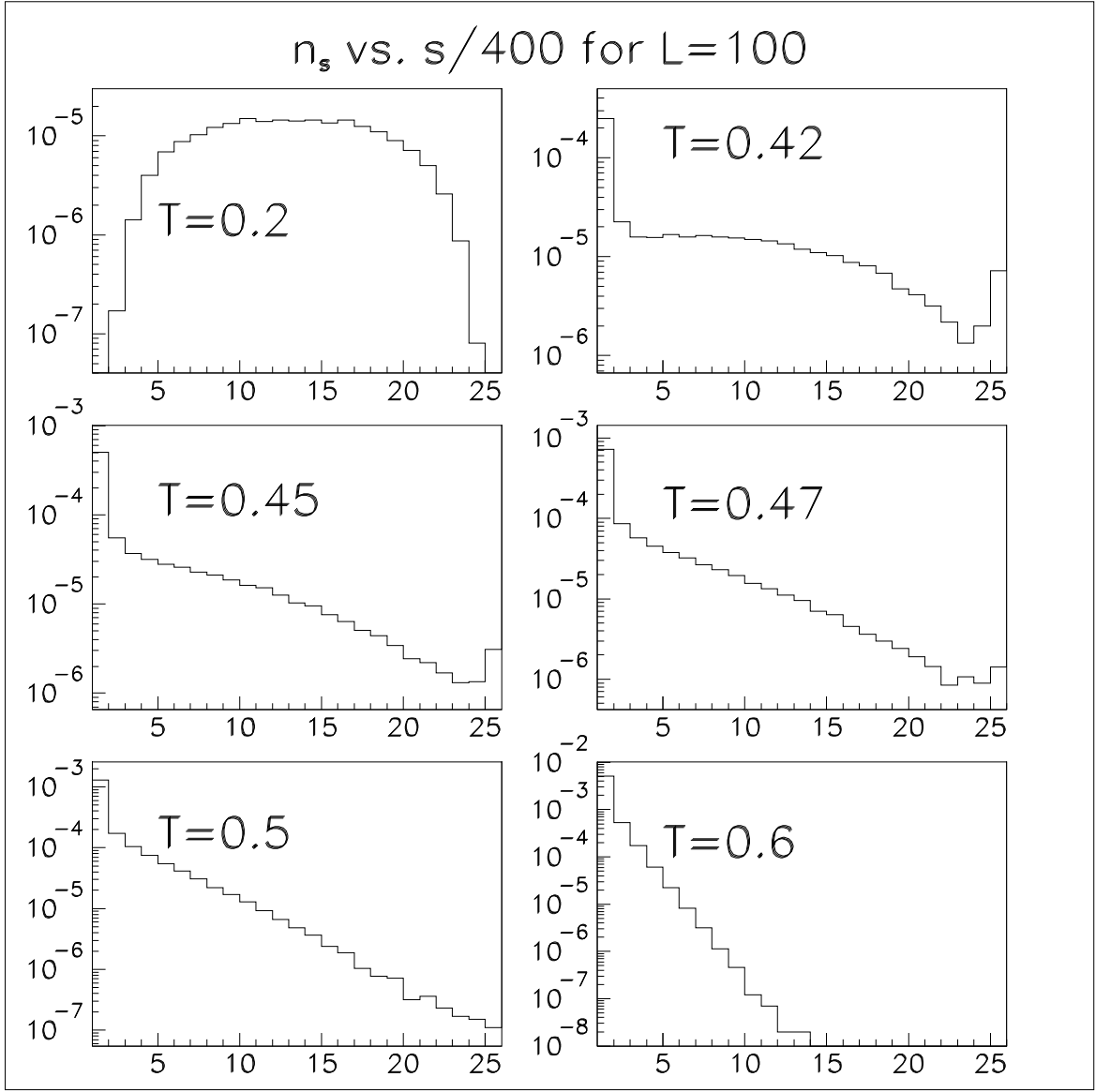 } }
\caption{Cluster distribution $n_s$ vs. cluster size $s$ for a system with
size $L=100$ at several temperatures $T$. Every bin is large 400 unities in 
cluster size. The percolation temperature for this system size is 
$T_p(L=100)\simeq 0.46$ then for all $T\geq 0.47$, even if the highest 
bin is not empty, there are only no percolating clusters. Let's note that for 
all $T$ above $T_p(L)$ it is $n_s\sim e^{-s}$ and that the distribution become 
symmetric for $T\rightarrow 0$.
}
\end{figure}

\begin{figure}
\mbox{ \epsfxsize=10cm \epsffile{ 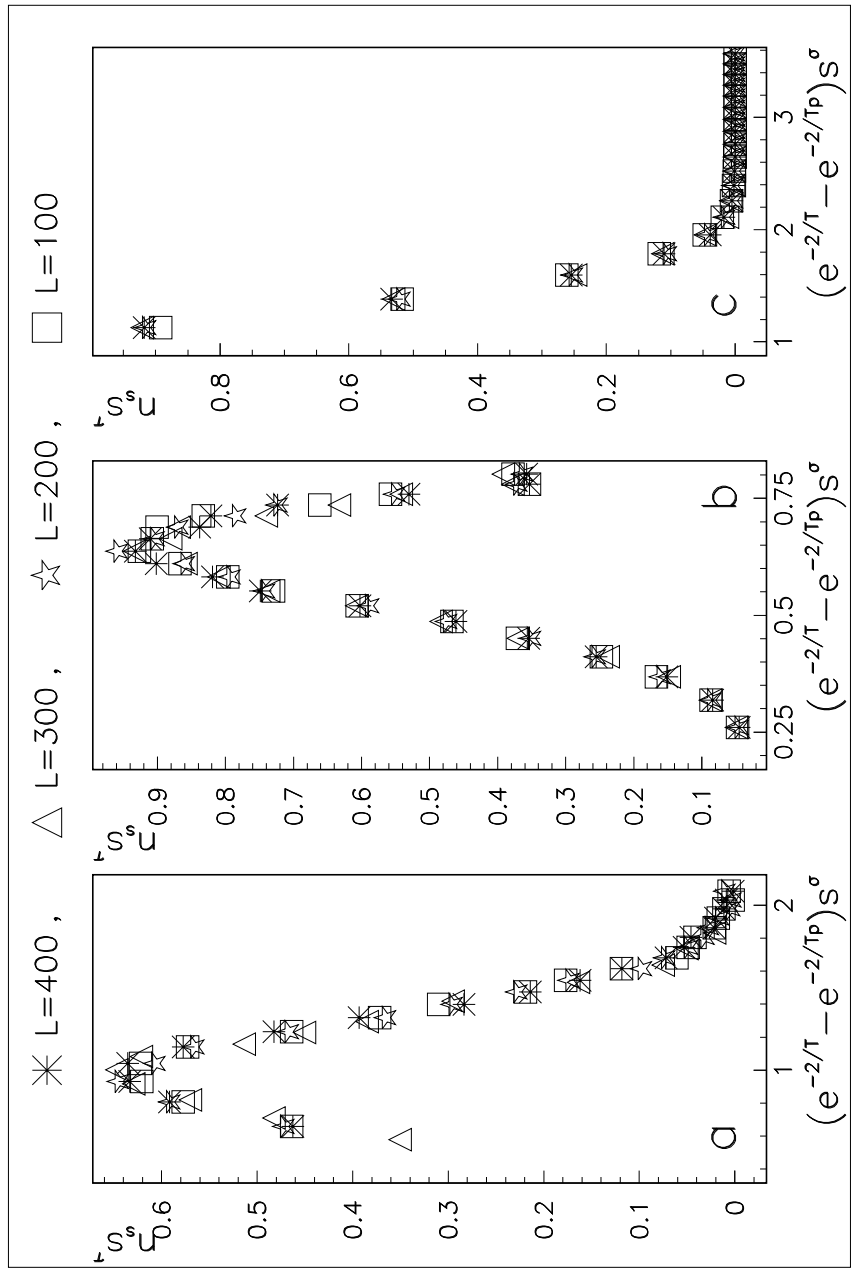 } }

\vspace{1.5cm}

\caption{a) Scaling for $n_s$ following assumption (\ref{scaling_ns}) 
with parameter $\tau=2.00\pm 0.01$ and $\sigma=0.50\pm 0.01$ for data
of systems with sizes $L=100,200,300,400$. Each set of data is chosen at a 
temperature near the corresponding $T_p(L)$. As consequence for each 
temperature the quantity $(e^{-2/kT}-e^{-2/T_p})L^{1/\nu}$ with $T_p=0$ 
and $\nu=1$ is equal to 2.084. Every point in the graph is an average over 500 
consecutive values of $s$ for $L=100$, 2000 for $L=200$, 4500 for $L=300$, 8000 
for $L=400$.
b) As in part a) but for $(e^{-2/T}-e^{-2/T_p})L^{1/\nu}=
0.823$ (below $T_p(L)$).
c) As in part a) but for $(e^{-2/T}-e^{-2/T_p})L^{1/\nu}=3.567$
(above $T_p(L)$).
}
\end{figure}

\begin{figure}
\mbox{ \epsfxsize=13cm \epsffile{ 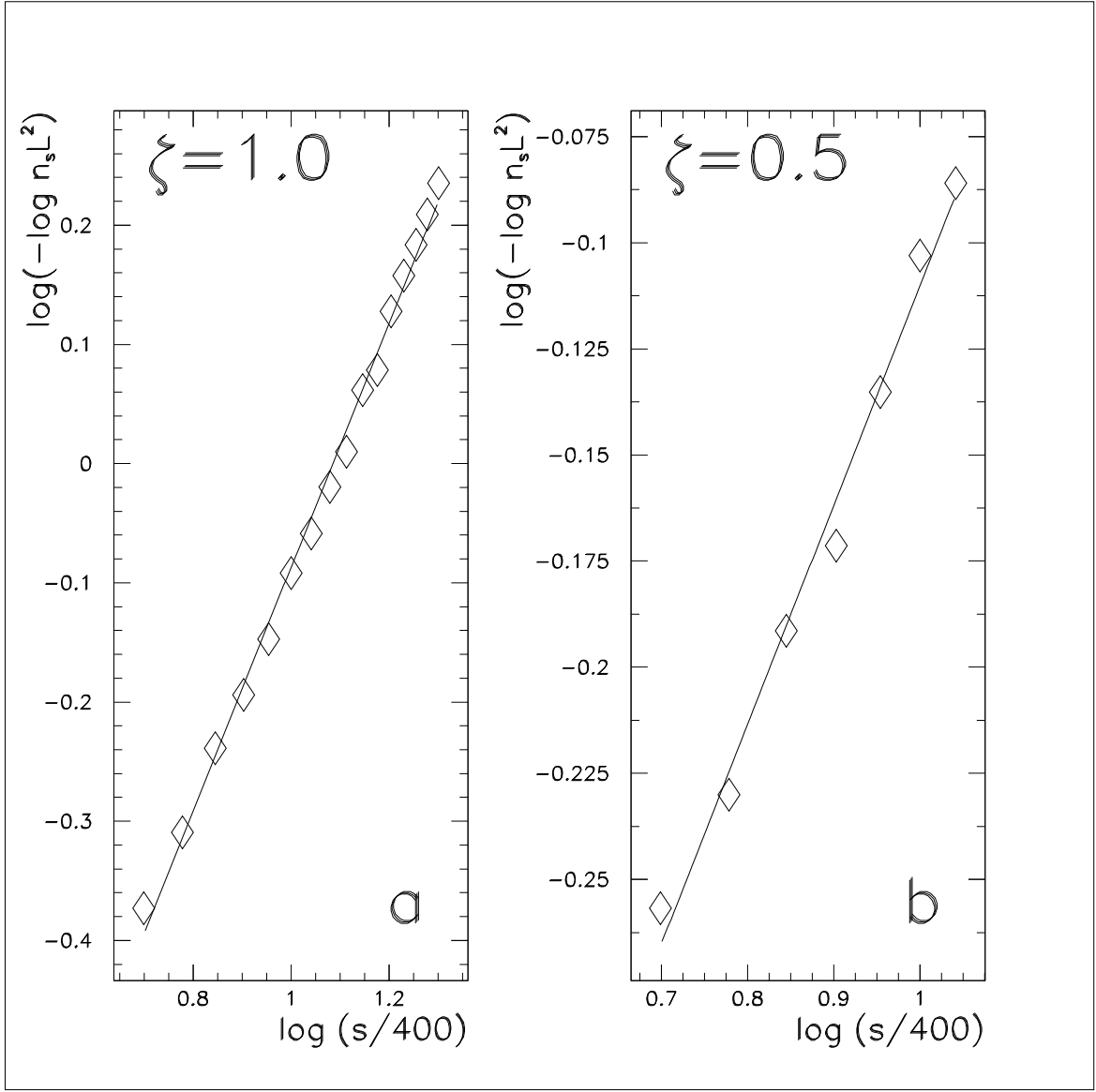 } }
\caption{Fit of $\log(-\log n_sL^2)$ vs. $\log (s/400)$ for a system with 
$L=100$ and $T_p(L)\simeq 0.46$: a) at $T\simeq 0.47>T_p(L)$ the slope is 
$\zeta\simeq 1$; b) at $T\simeq 0.45<T_p(L)$ the slope is $\zeta\simeq 1/2$.
}
\end{figure}

\begin{figure}
\mbox{ \epsfxsize=13cm \epsffile{ 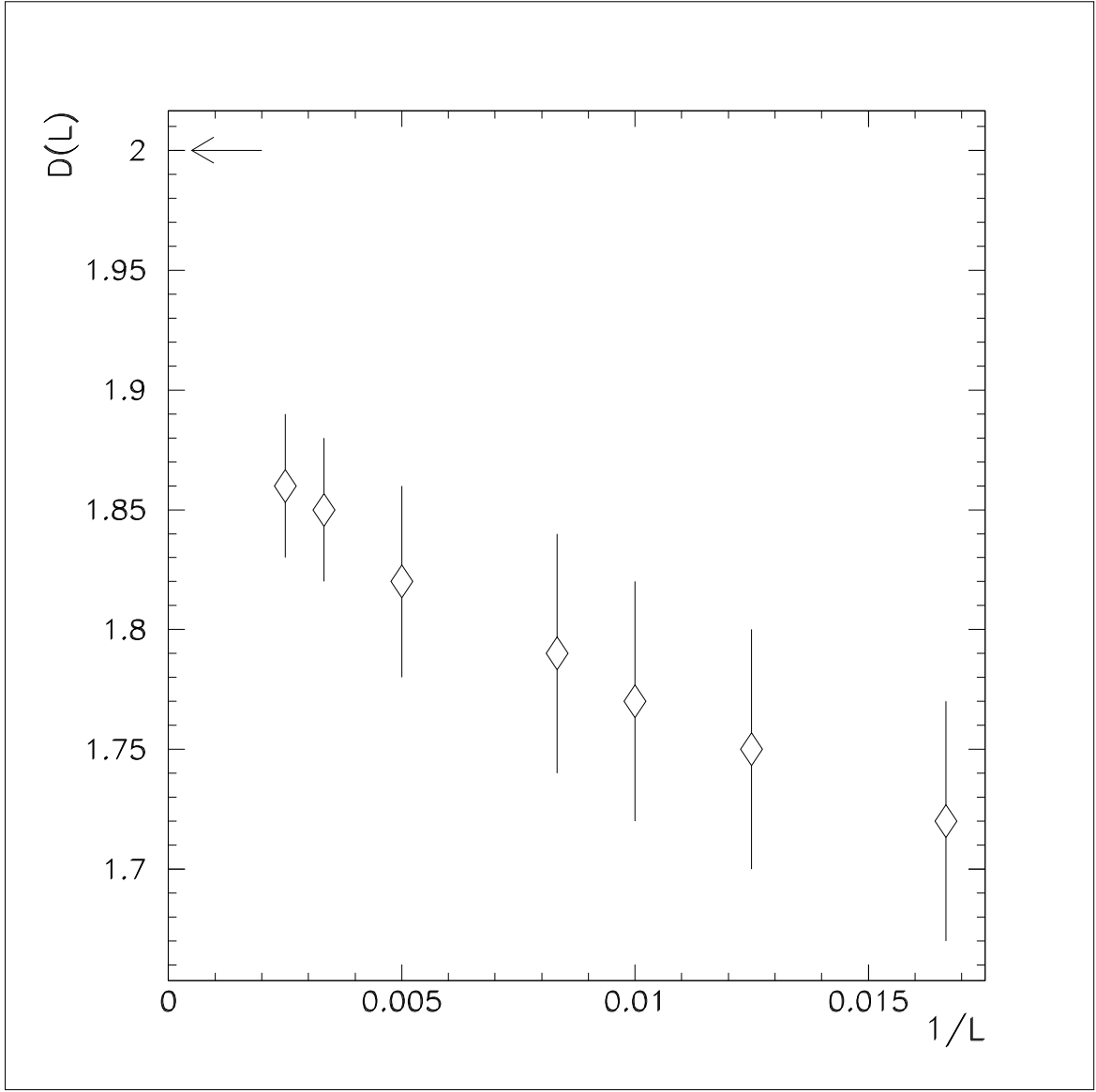 } }
\caption{Estimate of $D$ from definition (\ref{def_D}) vs. $1/L$ (see
Tab.1): at $T=0.517$ for $L=60$, at $T=0.481$ for $L=80$, at $T=0.450$ for 
$L=100$, at $T=0.437$ for $L=120$, at $T=0.389$ for $L=200$, at $T=0.360$ for 
$L=300$ and at $T=0.343$ for $L=400$. The error bars are probably
underestimated. The arrow heads for the asymptotic value of $D$.
}
\end{figure}
\end{center}

\end{document}